\documentclass[11pt,a4paper]{article}
\usepackage[T1]{fontenc}
\usepackage[utf8]{inputenc}
\usepackage{authblk}
\usepackage{graphicx}

\title{Ab initio calculations for non-strange and strange few-baryon systems}

\author{Winfried Leidemann}

\affil{Dipartimento di Fisica, Universit\`a di Trento, I-38123 Trento, Italy \\
              and INFN, TIFPA, I-38123 Trento, Italy \\
}

\begin{document}
  \maketitle

\begin{abstract}
Concerning the non-strange particle systems the low-energy excitation spectra of the three- and 
four-body helium isotopes are studied. Objects of the study are the astrophysical $S$-factor 
$S_{12}$ of the radiative proton deuteron capture $d(p,\gamma)^3$He and the width of the $^4$He 
isoscalar monopole resonance. 
Both observables are calculated using the Lorentz integral
transform (LIT) method. The LIT equations are solved via expansions of the LIT states on a
specifically modified hyperspherical harmonics (HH) basis. It is illustrated that at low 
energies such a modification allows to work with much higher LIT resolutions than with an 
unmodified HH basis. It is discussed that this opens up the possibility to determine astrophysical 
$S$-factors as well as the width of low-lying resonances with the LIT method. In the 
sector of strange baryon systems binding energies of the hypernucleus $^3_\Lambda$H are calculated using 
a nonsymmetrized HH basis. The results are compared with those calculated by various other 
groups with different methods. For all the considered non-strange and strange baryon systems it 
is shown that high-precision results are obtained. 

\end{abstract}

\maketitle
\section{Introduction}

Non-strange and strange few-baryon systems are particularly interesting particle systems
in the hadronic sector. On the one hand they serve for parametrization and test of potential 
models for nucleon-nucleon (NN), nuc-leon-hyperon (NY), hyperon-hyperon (YY) interactions and 
the various analogous three-body interactions. On the other hand they play an important
role in testing the quality of ab initio methods, for example in benchmark calculations.
In the present work the second of these two aspects is of relevance. In fact our aim is
to test the quality of two different ab initio techniques for specific physical questions
as is explained in greater detail in the following.

One of the tested ab initio methods is the LIT. The LIT approach is well-established~\cite{EfL07}
and allows to determine observables involving the many-body continuum without the necessity to 
calculate continuum wave functions. In the present work it is investigated to what extent 
specific features in the low-energy electromagnetic response of nuclei can be determined
with the LIT method. We consider two examples: (i)
the width of the $^4$He isoscalar monopole resonance $0^+$ and (ii)
the threshold cross section in $^3$He photodisintegration. The
inverse reaction of the latter, the radiative proton-deuteron capture,
is of relevance for the nucleosynthesis and usually parametrized via the astrophysical $S$-factor $S_{12}$.
As explained in section~\ref{secLIT} the crucial point for an exact description of the observables
mentioned above lies in the question whether a sufficiently high density of LIT states can be obtained in the 
low-energy region. In a rather recent LIT calculation based on HH expansions~\cite{BaB13}, where the $^4$He 
inelastic isoscalar monopole response function was computed with realistic nuclear forces, this aim could not
be achieved even though the HH basis was quite large. Therefore it was not possible to determine the width 
of the $0^+$ resonance in this calculation. In \cite{Lei15} it was then shown that the problem is 
due to the employed HH basis and that a somewhat modified many-body basis solves the problem of 
a too low low-energy density of LIT states. The modification consists in using 
for an $A$-body system, instead of an $A$-body HH basis, an $(A-1)$-body HH basis times 
a basis set for the relative motion of the $A$-$th$ particle with respect to the center of mass 
of the $(A-1)$-body system.  

For the strange particle systems responses to external probes have not yet been determined
in experiment. In fact the knowledge of such systems is still rather scarce. One source of experimental 
information are binding energies of hypernuclei. One of the future aims of our group are ab initio 
calculations of such binding energies with realistic forces. However, first we want to test the
reliability and the precision of the ab initio approach chosen by us. 
Different from the basis systems, which is used for the above mentioned LIT calculations, where
the various HH basis states have a well-defined permutational symmetry, we take
for the bound-state calculations of strange few-body baryon systems a nonsymmetrized HH (NSHH)
basis. The results are then compared with results coming from other ab initio approaches.
We switch from a symmetrized to a nonsymmetrized HH basis because
we are confident that calculations for $A$-body baryon systems with $A\ge6$ can be carried out with less 
computational effort. 

Since in the present work it is the aim to test the precision of various theoretical ab initio approaches 
we do not employ realistic interaction models, but use instead simpler 
potential models, which will be defined in the following sections. 

This work is organized as follows. In section~\ref{secLIT} the LIT method and the used many-body basis 
systems are briefly described. Furthermore, the LIT results for the above mentioned $S$-factor $S_{12}$ 
as well as for the $0^+$ resonance of $^4$He are discussed. In both cases results with the HH and the
new basis are compared. In section~\ref{secNSHH} it is described how
nonsymmetrized basis systems can nonetheless be used to determine ground states of systems
which obey a specific permutational symmetry. Subsequently the results for binding and $\Lambda$ separation 
energies of $^3_\Lambda$H are illustrated in comparison to results from other authors
with different ab initio few-body techniques. Finally, in section~4 a summary is given.
 
\section{The LIT method}
\label{secLIT}

Nuclear cross sections of inclusive reactions with electromagnetic probes are expressed
in terms of inclusive response functions, which contain the information
about the dynamics of the nucleus under investigation. Inclusive response
functions are in general of the following form
\begin{equation}
\label{response}
R(\omega) = \int df |\langle f| {\hat O} | 0\rangle|^2 \delta(E_f - E_0 - \omega) \,, 
\end{equation}
where $|0 \rangle$ and $|f\rangle$ are nuclear ground and final states, $E_0$ and $E_f$ are 
the corresponding eigenenergies and $\omega$ is the energy of the exchanged real (photoabsorption)
or virtual photon (electron scattering). Finally, ${\hat O}$ denotes the operator inducing the 
reaction.

A calculation of $R(\omega)$ can be become very difficult or even impossible for cases where
$|f\rangle$ is a many-body continuum state. However an explicit calculation of $|f\rangle$ can
be avoided by the use of the LIT, which is an integral transform defined as follows
\begin{equation}
\label{LIT}
L(\sigma) = \int d\omega \, {\frac{R(\omega)}{(\omega-\sigma_R)^2 + \sigma_I^2}} \, 
\end{equation}
with $\sigma = \sigma_R + i \sigma_I$. Due to the variable width of $2 \sigma_I$ of the Lorentzian kernel
the LIT is an integral transform with a controlled resolution. But it is important to realize that
in a given calculation one cannot
simply increase the resolution by choosing smaller and smaller $\sigma_I$ values. In fact one
has to make sure that the precision of the LIT calculation allows the choice of a smaller
$\sigma_I$ value. How this can be achieved becomes clearer in the discussion that follows next.

Since the aim is to determine the response function without the knowledge of the continuum wave
function it is useless to calculate the LIT via its definition of eq.~(\ref{response}). 
Fortunately, the LIT can be determined in an alternative way, namely by solving an equation, 
the LIT equation, given by  
\begin{equation} 
\label{eqLIT}
({\hat H}-E_0-\sigma) \, |\tilde\Psi(\sigma) = {\hat O}  | 0\rangle \,,
\end{equation}
where ${\hat H}$ is the nuclear Hamiltonian. The important feature of the solution
$\tilde\Psi(\sigma)$ is that it is a localized function. Therefore one can compute
$\tilde\Psi(\sigma)$ using bound-state methods. 

After having determined $\tilde\Psi(\sigma)$ one calculates the LIT from the following expression
\begin{equation}
L(\sigma) = \langle \tilde\Psi(\sigma) | \tilde\Psi(\sigma) \rangle \,.
\end{equation}
In order to obtain the response function $R(\omega)$ one has to invert the LIT.
Details about inversion methods are described in \cite{EfL07,Lei08}.

As already mentioned in the introduction an expansion on a complete many-body
basis is used for the solution of the LIT equation~(\ref{eqLIT}). First, the Hamiltonian
matrix for such a basis is determined, then, in a subsequent  diagonalization of this
matrix, $N$ eigenvalues $E_n$ and eigenstates $\phi_n$ (LIT states)
($n = 1,2,...,N$) are obtained, where $N$ is the dimension of the basis. The LIT can then be expressed in
terms of the energy eigenvalues and the LIT states. One obtains
\begin{equation}
\label{LIT_En}
 L(\sigma) = \sum_{n=1}^N {\frac {S_n}{(\sigma_R-(E_n-E_0))^2 + \sigma_I^2}} \,.
\end{equation}
with
\begin{equation}
S_n = |\langle \phi_n| {\hat O} | 0 \rangle |^2 \,. 
\end{equation}
Coming back to the question which resolution or in other words which value of $\sigma_I$ can be sustained
in a given LIT calculation it is already pointed out in the introduction that the density of LIT states 
plays a crucial rule. In fact the higher the density of LIT states the higher is also the resolution. This 
point will be better illustrated in section~\ref{subsec2_2}.

\subsection{Many-body basis systems}
\label{subsec2_1}

In most of the LIT applications the HH basis has been used for the expansions of
ground-state wave functions and LIT states of the considered $A$-body system.
This is mainly attributed to its property of being a complete $A$-body basis for localized states.

An HH basis has the following form
\begin{equation}
\label{HH}
HH_{[K]n}(\Omega_A,\rho_A) = {\cal Y}_{[K]}(\Omega_A) R_n(\rho_A) \, .
\end{equation}
It consists of an hyperangular part ${\cal Y}_{[K]}(\Omega_A)$ and a hyperradial part $R_n(\rho_A)$,
where $\Omega_A$ is a set of $3A-4$ hyperangles, $\rho_A$ denotes the hyperradius and $[K]$
stands for a set of hyperspherical quantum numbers. For the hyperradial basis functions 
Laguerre polynomials $L_n^{(\beta)}(\rho_A)$ times an exponential factor $\exp(-\rho_A/2b)$ are used, 
where $\beta$ and $b$ are a free parameters. In addition one can also introduce as a multiplicative
factor NN short-range correlation functions of Jastrow type, which can be purely central or
can also become spin and/or isospin dependent (see e.g.~\cite{EfL00}).

Usually the hyperangular states  ${\cal Y}_{[K]}$
are constructed with a well-defined permutational symmetry. With a complementary permutational
symmetry of the spin-isospin part of the nuclear wave function one then obtains an antisymmetric
basis. In case of $A=2$ the hyperangular basis functions reduce to the well-known spherical
harmonics $Y_{lm}$. A detailed description of the HH expansion technique is given e.g. in \cite{EfL07}. 

In the LIT applications of the present work a new basis $\Phi_{[K]nn'l}$ is employed in addition. It 
consists of a separation of the $A$-body basis in a $(A-1)$-part with HH basis functions 
$R_n(\rho_{A-1}) {\cal Y}_{[K]}(\Omega_{A-1})$ and a single-particle part with basis functions
$R_{n'}^{(2)}(r_A') Y_{lm}(\Omega_{r_A'})$:
\begin{equation}
\label{new}
\Phi_{[K]nn'l} =  {\cal Y}_{[K]}(\Omega_{A-1}) \, R_n(\rho_{A-1})\, R_{n'}^{(2)}(r_A') 
                  Y_{l(m)}(\Omega_{r_A'})
\end{equation}
with
${\bf r'}_A = {\bf r}_A - {\bf R}_{cm}^{(A-1)}$,
where ${\bf r}_A$ and ${\bf R}_{cm}^{(A-1)}$ 
is the position of the $A$-$th$ particle and the center of mass of the $(A-1)$-particle system, 
respectively. For $R_{n'}^{(2)}(r'_A)$ a similar expansion as for the hyperradial part is taken, 
namely a Laguerre polynomial  $L_{n'}^{(2)}(r_A')$ times an exponential factor $\exp(-r'_A/2b_A)$. 
Also in this case one may use the NN correlation functions, discussed above, in addition.

Of course, as for the HH basis given in eq.~(\ref{HH}), one has to multiply the basis functions of
eq.~(\ref{new}) with appropriate $A$-body spin-isospin wave functions and then one has to care for an 
antisymmetric state by making a proper antisymmetrization of the basis states.

\subsection{Photodisintegration of $^3$He and astrophysical $S$-factor $S_{12}$}
\label{subsec2_2}
\vskip 1cm
\begin{figure}[htb]
\centerline{\includegraphics[width=0.75\columnwidth]{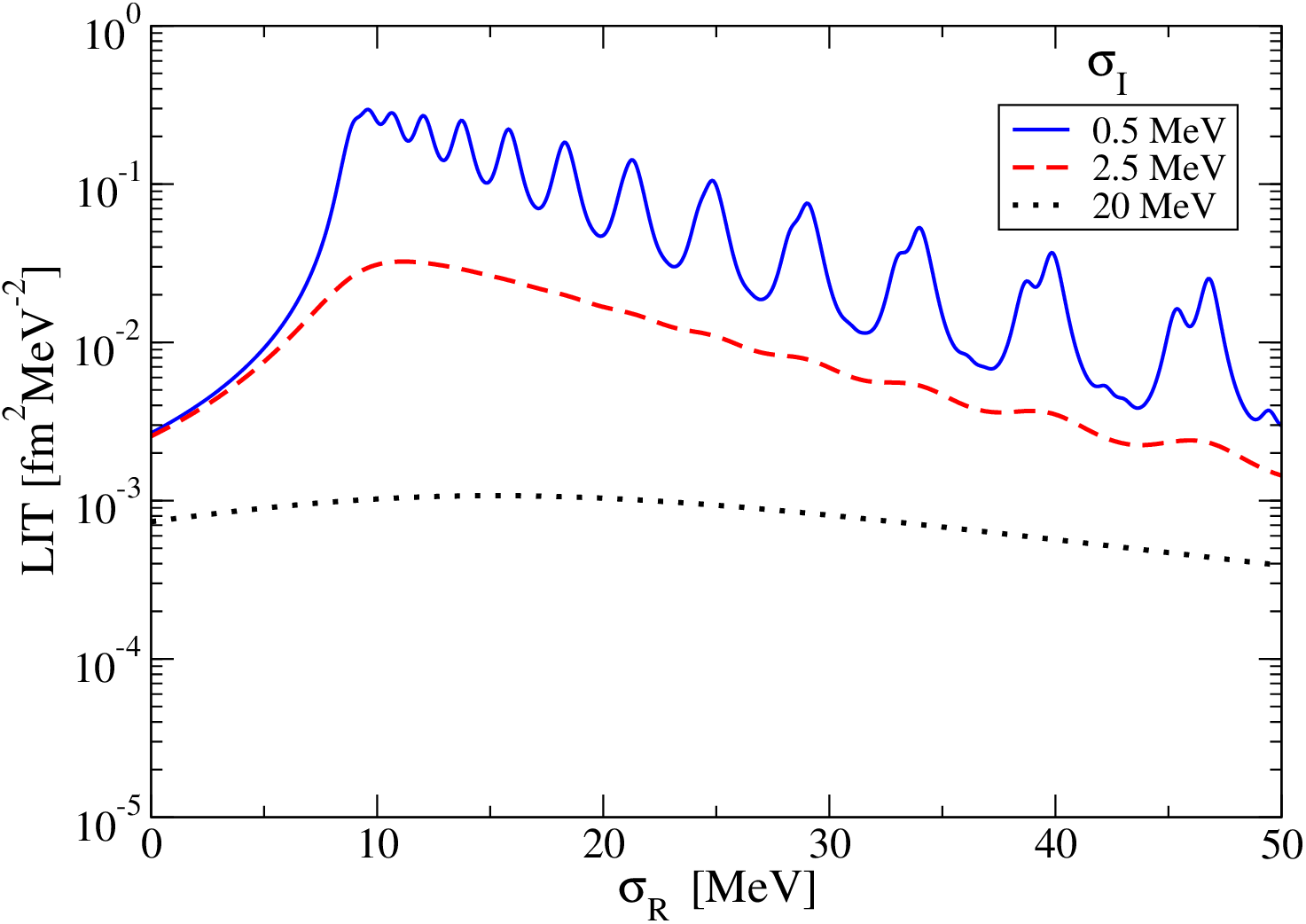}}
\caption{LIT of the $^3$He response function $R_{\rm E1}(\omega)$ using an HH basis with 31 hyperradial ($b = 0.3$ fm) and 30
hyperangular basis states.}
\end{figure}

As pointed out in the introduction, in this work the $S$-factor of the reaction $d(p,\gamma)^3$He is 
determined via the inverse reaction, the $^3$He photodisintegration, then time reversal invariance is 
applied to obtain $S_{12}$.

We take the unretarded dipole approximation for the calculation of the $^3$He 
photodisintegration cross section, which is given by
\begin{equation}
\label{xsec}
\sigma_{\rm E1}(\omega) = 4 \pi^2 \alpha \omega R_{\rm E1}(\omega) \,,
\end{equation}
where $\alpha$ is the fine structure constant and $R_{\rm E1}(\omega)$ denotes
the dipole response function. In this case
the components ${\hat O}$ and $|0\rangle$ of eq.~(\ref{response}) become equal to $D_z$, the third 
component of the nuclear dipole operator ${\bf D}$, and the $^3$He ground-state wave function,
respectively. 

In order to determine the $S$-factor $S_{12}$ one only needs to take into account the low-energy $^3$He 
photoabsorption cross section, which is exclusively due to the two-body breakup channel 
$^3$He$ + \gamma \rightarrow p + d$. Since the $pd$ channel has isospin $T=1/2$, 
only the $T=1/2$ channel is considered for the LIT equation~(\ref{eqLIT}). As NN potential
the MT-I/III potential~\cite{MaT69} is employed. To speed up the convergence of the expansions,
both with HH basis and new basis, the already mentioned central NN short-range correlation functions are 
taken into account in addition.

First LIT results with a three-body HH basis are considered. As mentioned in section~\ref{subsec2_1} 
the hyperradial basis functions contain Laguerre polynomials $L_n^{(\beta)}$, here $\beta=5$ is taken.
In fig.~1 the LIT for the case with 30 hyperangular and 31 hyperradial states ($b = 0.3$ fm) is shown. 
One sees that a smooth LIT is obtained with $\sigma_I =20$ MeV, while with $\sigma_I = 2.5$ MeV 
the contributions of single LIT states becomes visible at higher energies. Such contributions due to
single LIT states become even the dominant feature for $\sigma_I = 0.5$ MeV. This is a clear sign
that the LIT-state density is too low to  support a resolution
with a $\sigma_I$ value of 0.5 MeV. In fact the resolution of strength encoded in the LIT depends on 
the relative distance $\Delta E$ of two neighbouring LIT states. Structures with a 
width smaller than $\Delta E$ can hardly be resolved by an inversion of the LIT. In other words the 
higher the density of LIT states the finer the details that can be resolved inverting the LIT. 
\vskip 1cm
\begin{figure}[htb]
\centerline{\includegraphics[width=0.75\columnwidth]{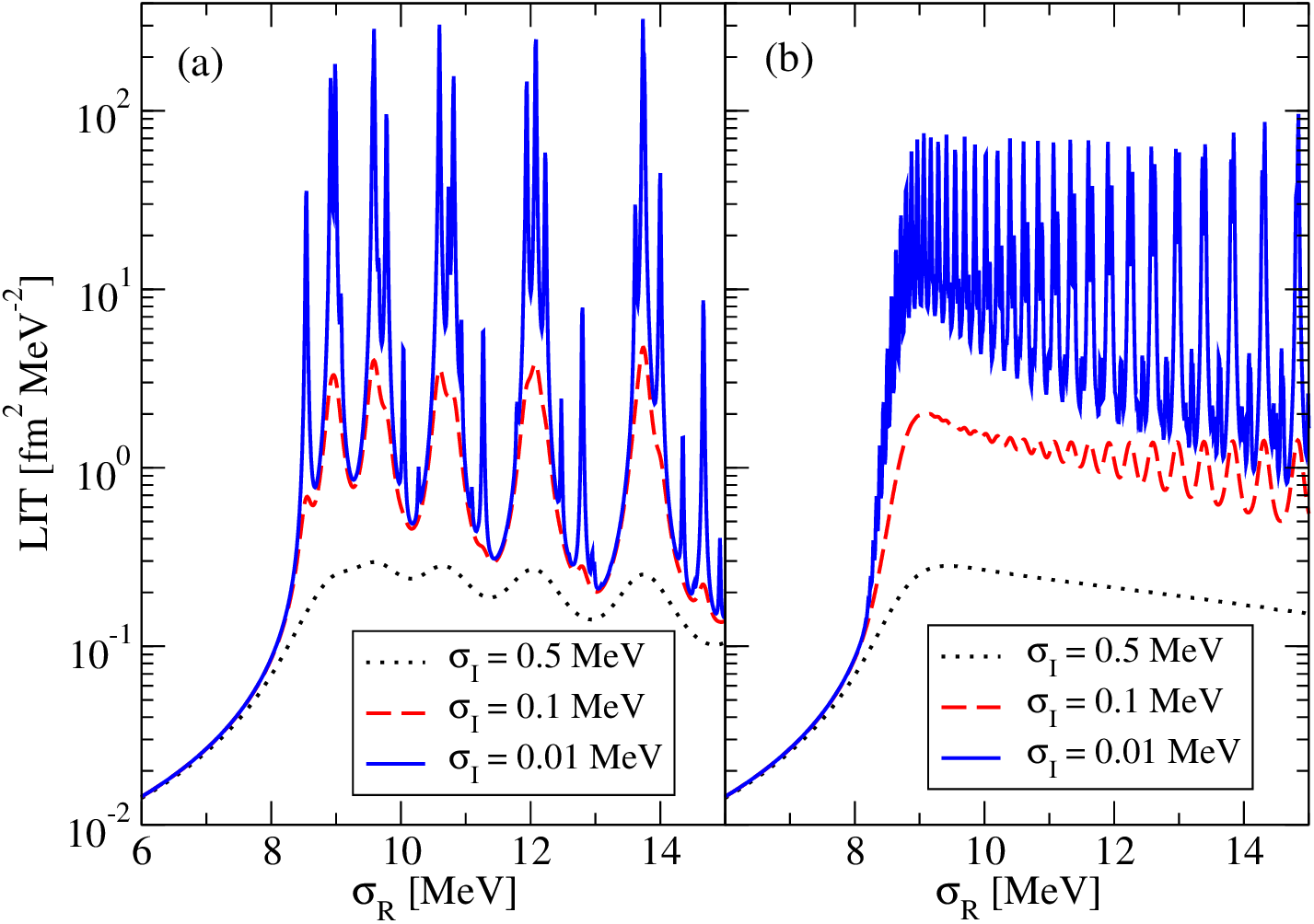}}
\caption{LIT of the $^3$He response function $R_{\rm E1}(\omega)$ using an HH basis as in fig.~1 (a)
and with 76 hyperradial ($b = 1$ fm) and 40 hyperangular basis states (b).}
\end{figure}
In order to enhance the LIT-state density one can increase the number of HH basis states taking more  
hyperangular and/or hyperradial states. In addition one can use a larger spatial extension of the basis 
by taking a greater value for the hyperradial parameter $b$, which then leads to a shift of LIT states 
towards lower energies. The resulting effects on the LIT are discussed in greater detail
in \cite{DeE17}. Here, in fig.~2, we only compare the LITs of fig.~1 at low energies with those obtained 
with an HH basis of 40 hyperangular and 76 hyperradial states with $b = 1$ fm. From the LIT results  
with $\sigma_I = 0.01$ MeV it is readily seen that the density of LIT states becomes much 
larger with the increased HH basis. Accordingly one finds for the lower resolutions of $\sigma_I$ equal 
to 0.1 and 0.5 MeV much smoother LIT results with the increased HH basis. However, one also notes a 
very important point:
even in fig.~2b there is not a single LIT state below the three-body breakup threshold at about 8 MeV 
($^3$He binding energy with MT potential). Thus the information about the response function is only
rather scarce in the energy range between the two-body breakup threshold at about 5.8 MeV and the 
three-body breakup threshold. Also with regard to the results following in section~\ref{subsec2_3} and to 
those of~\cite{BaB13} one may conclude that for an HH basis a systematic increase of the LIT-state density 
cannot easily be achieved in an energy region where only two-body breakup channels are open. Therefore 
the HH basis is not very suitable to obtain precise results for such cases, as for example the present one 
of low-energy $^3$He photodisintegration.

\vskip 1cm
\begin{figure}[htb]
\centerline{\includegraphics[width=0.75\columnwidth]{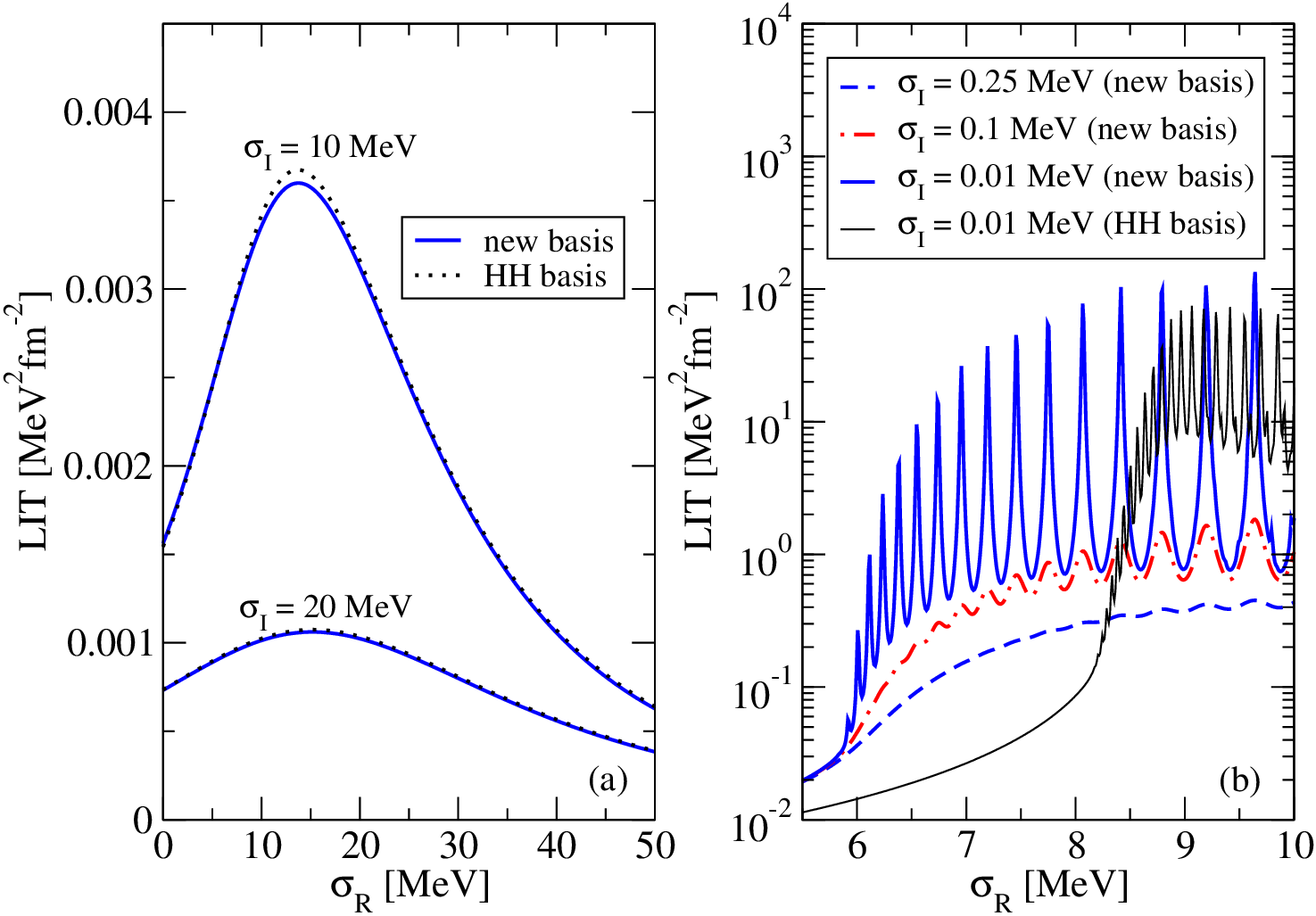}}
\caption{LIT of the $^3$He response function $R_{\rm E1}(\omega)$ using an HH basis as in fig.~2b (a)
and new basis as described in the text (b).}
\end{figure}

Now we turn to the results with the new basis described in section~\ref{subsec2_1}. The $(A-1)$-basis
corresponds in the present three-body case to a two-body basis with basis states
$Y_{lm}(\Omega_r) \, L_{n'}^{(2)}(r) \, \exp(-r/b_2)$ with ${\bf r} = {\bf r}_2 - {\bf r}_1$, where
${\bf r}_i$ is the position of the $i$-$th$ particle. A basis is used with 25 and 80 radial states for 
the two-body and single-particle basis, respectively ($b_2=0.75$ fm, $b_3=0.5$ fm). Since in the present
case only the low-energy part of the response function is relevant, it is sufficient to take into account
only s-states for the two-body basis. 

In fig.~3 LITs resulting from the increased HH basis of fig.~2b and those obtained with the 
new basis are shown. Figure~3a illustrates that both results are very similar for $\sigma_I=20$ MeV, while 
with $\sigma_I=10$ MeV one finds some differences in the region of the maximum. For a much smaller 
$\sigma_I$ of 0.01 MeV, shown in fig.~3b, strong differences become evident. In fact, only with the new 
basis LIT states are present right above the two-body breakup threshold at about 5.8 MeV. 
Moreover these states have a rather high density. More details of the LIT calculation with the new basis, 
like for example the convergence behaviour with respect to the two-body and single-particle basis systems, 
are discussed in \cite{DeE17}.

The results presented in fig.~3 show that the use of a proper many-body basis can become important for
specific questions. There are two conditions which should be considered: (i) is the density of LIT states
sufficiently high in order to extract specific structures in the response function and (ii) is the 
LIT-state density sufficiently regular in order to work with a single $\sigma_I$ value. If the
second condition is not fulfilled one should take in a region of lower LIT-state density a different,
more suitable, value for $\sigma_I$, otherwise one risks to misplace strength in the inversion. 
In fact in~\cite{DeE17} quite a number of different $\sigma_I$ values were used in order to take into 
account a lower LIT-state density with growing energy. 

For the LIT with the present HH basis one can conclude that the completely missing LIT states in the 
two-body breakup region do not only prevent to resolve the correct threshold behaviour of 
$R_{\rm E1}(\omega)$, but that one would also obtain an overestimation 
of the peak height of the response function if for the inversion one uses a LIT with a $\sigma_I$ value
much smaller than 20 MeV. As illustrated in~\cite{DeE17} (see fig.~7 therein) one can still obtain a 
rather reasonable inversion result with $\sigma_I=20$ MeV using the standard inversion method, described
in~\cite{EfL07}, where the correct threshold behaviour of the response function is implemented.

Concerning the LIT results with the new basis one can certainly say that the low-energy density of LIT 
states is quite high and that the pattern is very regular. Thus one may expect that the low-energy response,
and thus the astrophysical $S$-factor $S_{12}$, can be determined very precisely by the inversion.  
In fig.~4 we show the result for $S_{12}$ obtained with the new basis in comparison to a calculation 
of $S_{12}$ with explicit wave functions for the proton-deuteron continuum states
(for details of the continuum state calculation see \cite{DeE17}).
As anticipated one observes an excellent agreement between both results. Note that the figure also 
contains an error estimate due to the LIT inversion, also here we refer to \cite{DeE17} for a more 
detailed description of the error estimate.   

\vskip 1.1cm
\begin{figure}[htb]
\centerline{\includegraphics[width=0.75\columnwidth]{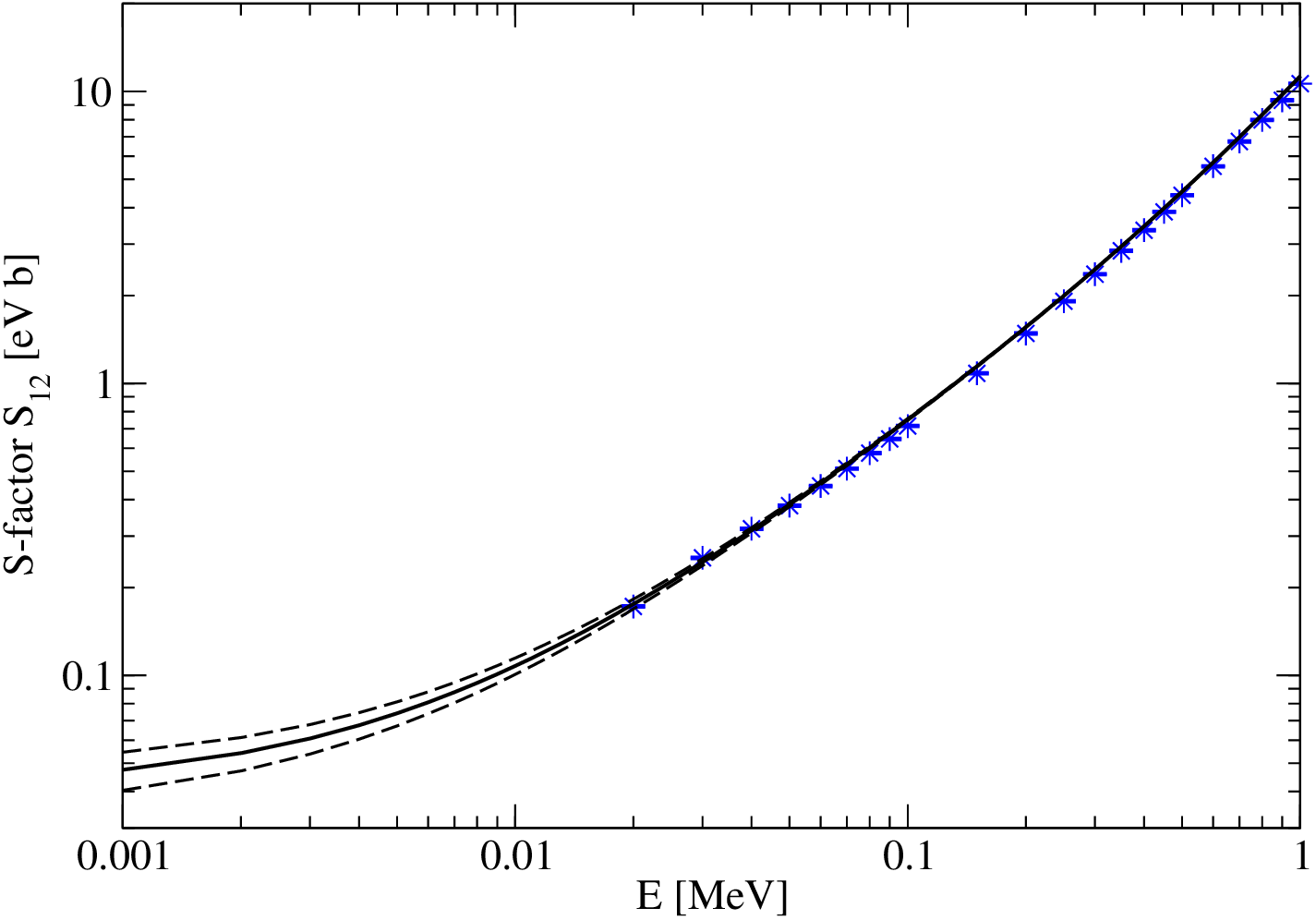}}
\caption{Astrophysical S-factor $S_{12}$ calculated with the LIT method (full curve) with additional
error estimate due to inversion (dashed curves); in addition results of a calculation with explicit 
$pd$ continuum wave functions (stars).}
\end{figure}

\subsection{The $^4$He isoscalar monopole resonance}
\label{subsec2_3}
The isoscalar monopole resonance $0^+$ of the $\alpha$-particle leaves a strong signal in inclusive 
inelastic electron scattering experiments~\cite{Wa70,Fr65,Ko83}. The corresponding transition form 
factor was studied in a LIT calculation, where an HH basis and modern realistic forces were 
used~\cite{BaB13}. A rather strong potential model dependence was found, but the experimental data were 
overestimated quite a bit. The 
present work, however, is not devoted to determine the strength of the transition form factor, but 
rather to a different aspect of the resonance, namely its rather small width of 270(50) keV as
determined in the $^4$He$(e,e')$ experiments mentioned above.  In~\cite{BaB13} 
this question could not be addressed because the density of LIT states was not sufficiently high in
the region of the $0^+$ resonance, which is located closely above the lowest $^4$He  two-body 
breakup threshold.

The isoscalar monopole response function $R_{\rm C0}(q,\omega)$ depends on energy transfer $\omega$ and
momentum transfer $q$ mediated in electron scattering by the exchanged virtual photon.
Thus the corresponding transition operator ${\hat O}$ of eq.~(\ref{response}) becomes $q$-dependent:
\begin{equation}\label{monopole}
 {\hat O}(q) =\frac{G_E^s(q^2)}{2} \sum_{i=1}^A \, j_0(q r_i)\,.
\end{equation}
In the equation above $G_E^s(q^2)$ is the nucleon isoscalar electric form factor,
${\bf r}_i$ is the position of nucleon $i$,
and  $j_0$ is the spherical Bessel function of 0$^{th}$ order.

For the present study the LIT of $R_{\rm C0}(q,\omega)$ is taken at $q=300$ MeV/c, a $q$ value, which 
lies in the 
momentum transfer range of maximal strength of the $0^+$ transition form factor. Here we consider results 
with an HH basis for the four-body system and in addition the new  basis as described in 
section~\ref{subsec2_1} (three-body HH basis plus single-particle basis). For details of the used
basis states we refer to~\cite{Lei15}. As NN potential model the central TN potential is taken, 
it has been used in the very first LIT applications for the $\alpha$-particle 
(see for example~\cite{EfL07}). Like in the previous case in section~\ref{subsec2_2}
central NN correlation functions are used in order to accelerate the convergence of HH and new basis.

In fig.~5 the LIT results for the response function  are shown. For the 
HH basis in fig.~5a one sees a very similar picture as in fig.~2a.
There is only one essential difference, in fig.~2a there are no LIT states below the many-body
breakup threshold, whereas in fig.~5a one finds just one LIT state below the many-body breakup threshold
at about 30 MeV (note $^4$He binding energy with TN potential is 31.4 MeV).
The isolated low-energy LIT state for the $^4$He case is due to the $0^+$ resonance. It is evident
that with a single LIT state it is impossible to determine a resonance width. In the already mentioned
LIT calculation for $R_{\rm C0}(q,\omega)$ with realistic nuclear forces of~\cite{BaB13} the situation was 
somewhat better, but the LIT-state density could not be systematically improved in order to have
sufficient information to compute the $0^+$ resonance width. 

As illustrated in fig.~5b the situation is much better in case of the new basis. It is interesting to study 
the results with the various $\sigma_I$ values. With a low resolution of $\sigma_I= 5$ MeV one does not 
realize that there is a resonance (note the logarithmic scale). If one increases the resolution using smaller 
$\sigma_I$ values the resonance becomes more and more distinct from the background. For the highest
resolution of $\sigma_I=$ 0.01 MeV one sees that there are various LIT states in the region
of the resonance at about 26 MeV. In fact in~\cite{Lei15} it was possible to determine
the width using besides the results of fig.~5b also LIT results with a different basis size (for 
details see~\cite{Lei15}). The obtained width of 180(70) keV agrees quite well with the experimental 
value of 270(50) keV.    

\vskip 0.75cm
\begin{figure}[htb]
\centerline{\includegraphics[width=0.75\columnwidth]{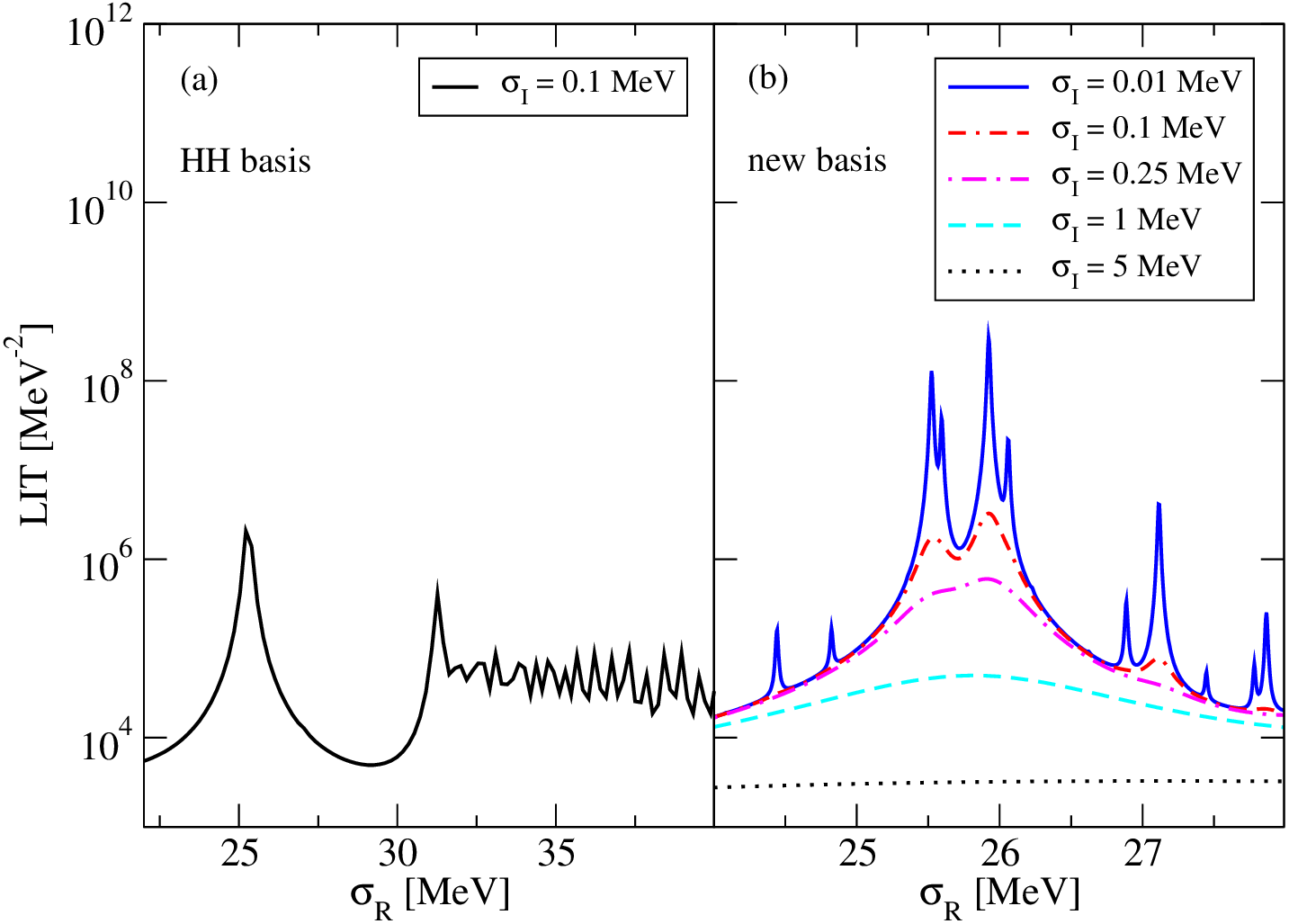}}
\caption{LIT of the $^4$He response function $R_{\rm C0}(q,\omega)$ at $q=300$ MeV/c using a four-body
 HH basis (a) and the new basis $\Phi_{[K]nn'l}$ of eq.~(\ref{new}) (b) (a detailed description of
the considered basis states is given in~\cite{Lei15}).}
\end{figure}

\section{Strange baryon systems}
\label{secNSHH}
In the benchmark calculation of~\cite{FeB17} bound baryon systems from three up to five particles are 
considered, where one of the baryons is the $\Lambda$ hyperon which has strangeness $S=-1$. As already mentioned 
in the introduction the experimental information about the YN interaction is still rather scarce. On the 
other hand, our present aim is not yet a realistic calculation of the binding energy of hypernuclei,
but rather a check of the precision of the ab initio method used by us. Therefore in~\cite{FeB17}
calculations with non-fully realistic interactions models are made. 
Before coming to some of these results in section~\ref{subsec3_2}, first, a short description of the ab 
initio method of our choice is given in the following section.

\subsection{The nonsymmetrized hyperspherical harmonics (NSHH) expansion}
The NSHH expansion relies on the HH expansion given in eq.~(\ref{HH}),
but the hyperspherical functions ${\cal Y}_{[K]}(\Omega_A)$ are not constructed with
any permutational symmetry. Also the spin-isospin part of a hypernuclear basis state is taken
without imposing a permutational symmetry. On the other hand it is clear that a hypernuclear
wave function has to be antisymmetric under the exchange of two identical fermions. 
At this point it is helpful to consider the Casimir operator ${\hat C}$ of the particle system. Taking 
an $A$-body baryon system with $N$ nucleons ($n=1,2,...,N$) and L $\Lambda$ hyperons 
($n=N+1,N+2,...,N+L=A)$ one has
\begin{equation}
{\hat C} = {\hat C}_N + {\hat C}_\Lambda = \sum_{j>i=1}^N {\hat P}_{ij} + \sum_{j>i=N+1}^A {\hat P}_{ij} \,,
\end{equation}
where the operator ${\hat P}_{ij}$ exchanges particles $i$ and $j$.
The eigenvalues $\lambda_{[I]}$ of the Casimir operator depend on the specific permutational symmetry
of the eigen functions.

In our case 
with just one $\Lambda$ particle only the permutational symmetry of the nucleons is relevant. One has the 
lowest eigenvalue for the antisymmetric case ($\lambda_{[A]}=-N(N-1)/2)$ and the highest for the 
symmetric case ($\lambda_{[S]}=+N(N-1)/2)$. Thus diagonalizing the Hamiltonian matrix for an NSHH basis, 
one can find out the symmetry of a given eigenstate by calculating the corresponding eigenvalue 
$\lambda_{[I]}$ applying the Casimir operator.
This is the strategy which has been put forward in~\cite{GaK11}. If, however, the
NSHH basis is very large and one wants to find the lowest state being antisymmetric for the nucleonic part, 
which in general
is not the lowest energy state, it is more convenient concerning the computational resources to apply 
the strategy of~\cite{DeB13}. In fact in case of large basis systems it is not advisable to perform
a complete diagonalization of $H$. It is much better to use the Lanczos technique, which saves 
computational resources and leads to a fast determination of the lowest energy state. In order to
bring the lowest antisymmetric state for the nucleonic part to the lowest overall state the following 
fictitious Hamiltonian ${\hat H}'$ has been introduced in~\cite{DeB13}:
\begin{equation}
{\hat H}' = {\hat H} + \gamma {\hat C}_N \,.
\end{equation} 
Thus, for a sufficiently large $\gamma$, such a lowest antisymmetric state will become the absolute ground 
state of the particle system with ground-state energy $E_0$. Therefore, taking ${\hat H}'$ instead of 
${\hat H}$, one can apply the Lanczos technique 
to find the proper ground state for an $A$-baryon system with $A-1$ nucleons and one $\Lambda$ hyperon. 
In order to have the correct bound-state energy one needs to correct $E_0$ only by $\gamma N(N-1)/2$.

\subsection{The  hypernucleus $^3_\Lambda$H}
\label{subsec3_2}
Here the benchmark results for one of the strange baryon systems discussed in~\cite{FeB17} are
illustrated, namely those for $^3_\Lambda$H. Two different potential sets were used in these calculations: 
(i) the AV4' 
NN potential~\cite{WiP02} together with the Bodmer-Usmani YN potential~\cite{BoU88} and (ii) the AV8' 
NN potential~\cite{WiP02} together with a parametrization~\cite{HiO14} of the meson-theoretical NSC97f YN 
potential~\cite{RiS99}. In order to accelerate the NSHH expansion an effective interaction is used as
described in~\cite{BaL00}.
Besides the NSHH results new results with the auxiliary field diffusion 
Monte Carlo (AFDMC) technique~\cite{LoG13} have been obtained. Also results due to the Faddeev
approach (FY) and due to the Gaussian expansion method~\cite{HiO14} (GEM)
are included in~\cite{FeB17}.

\begin{table}
\caption{\label{tab1}Binding energy BE and $\Lambda$ separation energy $S_\Lambda$ for 
$^3_\Lambda$H.}
\begin{center}
\begin{tabular}{llllll}
\hline
Potential model & Energy & NSHH & AFDMC & FY&GEM \\
\hline
AV4'+Bodmer-Usmani & BE & 2.530(3) & 2.42(6) & 2.537(1) & -\\
 & $S_\Lambda$ & 0.290(3) & 0.18(6) & 0.292(1) & - \\
\hline
AV8'+NSC97f & BE & 2.41(2) & - & 2.415(1) & -  \\
 & $S_\Lambda$ & 0.17(2) & - & 0.189(1)  & 0.19(1)\\
\hline
\end{tabular}
\end{center}
\end{table}

In table~1 the binding energy and the $\Lambda$ separation energy of $^3_\Lambda$H are listed
for the two different potential models defined above. One observes a rather good agreement between the 
various methods. Only the AFDMC results are a bit different, but this is not a real surprise since
the AFDMC is an ab initio method more suitable for systems with more than three particles and with some
preference for closed shell nuclei. In fact further calculations discussed in~\cite{FeB17} show that 
the comparison of AFDMC results with those of the other ab initio methods become decisively better for 
$^4_\Lambda$H and $^5_\Lambda$He. 

With the results of table~1 and the further ones given in~\cite{FeB17} one can conclude 
that the NSHH method is very well suited to give precise results for observables of hypernuclei.
A further benchmark with the AFDMC method, where also three-body interactions are taken into account,
will be published in the near future. Thus, more ambitious calculations with more realistic interaction
models can be tackled with the NSHH method in future.

\section{Summary}
The purpose of this work is twofold. Firstly, it is a check of the applicability of the LIT method
for a precise determination of specific details in nuclear low-energy cross sections that are
induced by external electromagnetic probes. To this end the reactions $^3$He$(\gamma)$ and 
$^4$He$(e,e')$ have been considered. The $^3$He photodisintegration has been calculated in order
to obtain the astrophysical $S$-factor $S_{12}$ of its inverse reaction, i.e. $d(p,\gamma)^3$He,
by applying time reversal invariance. Thus the actual aim has been a precise determination of $S_{12}$
via the LIT method. Comparing to results of a calculation with explicit proton-deuteron continuum
wave functions it has been shown that the LIT leads to excellent results for $S_{12}$. The calculation
has been carried out with a simple central NN potential, but the real importance of the calculation
does not lie in a realistic calculation of $S_{12}$, more important is the fact that the LIT method
could serve to calculate astrophysical $S$-factors of reactions involving more than three nucleons.
Also the inclusive electrodisintegration of $^4$He has been computed here with a central NN interaction.
Again, the essential aim has not been to obtain realistic results, but to test the possibility
to determine the width of a narrow resonance with the LIT method. In fact the $^4$He continuum exhibits a 
rather narrow resonance, the so-called isoscalar monopole resonance $0^+$. Therefore the LIT calculation 
has been performed for the $^4$He isoscalar monopole response function $R_{\rm C0}(q,\omega)$. It has been 
shown that the resonance width can be determined with the LIT method and that a value of 180(70) keV is
obtained. This agrees quite well with the experimental result of 270(50) keV. Thus, there is cause for hope
that the $0^+$ resonance width can be determined with the LIT method also for the case of a realistic 
nuclear force.

It has been pointed out that the decisive point for precise determinations of $S$-factor 
$S_{12}$ and of $^4$He $0^+$ resonance width is a sufficiently high density of LIT states. Unfortunately,
this seems to be very difficult to achieve if one uses as $A$-body basis an HH $A$-body basis.
It is much better to take an HH hybrid basis consisting of an $(A-1)$-body HH basis and an additional 
single-particle basis for the $A-th$ nucleon. It has been illustrated that with such a basis it is possible to
systematically increase the density of LIT states in the two-body breakup region at low-energies.

The second purpose of the present work has been a benchmark calculation for $^3_\Lambda$H.
The ab initio method of our choice, expansion of the hypernuclear ground state on a nonsymmetrized
HH basis, has been discussed, in particular, how such a basis can serve to obtain a ground state
with a proper permutational symmetry. Two different potential model sets have been employed for NN 
and N$\Lambda$ interactions. In comparison to results from other ab initio approaches it has been
found that the present calculation leads to reliable results for the $^3_\Lambda$H binding energy and 
the corresponding $\Lambda$ separation energy.

Finally, it should be mentioned that more details of the various calculations are given in
\cite{Lei15,DeE17,FeB17}.

\end{document}